\documentclass[onecolumn,showpacs]{revtex4}

\usepackage{graphicx}
\usepackage{dcolumn}
\usepackage{bm}

\input epsf

\begin{document}

\title{\Large Collapsing Inhomogeneous Dust Fluid in the Background of Dark Energy}

\author{\bf~Tanwi~Bandyopadhyay\footnote{b$_{-}$tanwi@yahoo.com}~and~Subenoy
Chakraborty\footnote{subenoyc@yahoo.co.in}}

\affiliation{Department of Mathematics, Jadavpur University,
Calcutta-32, India.}

\date{\today}

\begin{abstract}
In the present work, gravitational collapse of an inhomogeneous
spherical star model, consisting of inhomogeneous dust fluid
(dark matter) in the background of dark energy is considered. The
collapsing process is examined first separately for both dark
matter and dark energy and then under the combined effect of dark
matter and dark energy with or without interaction. The dark
energy is considered in the form of perfect fluid and both
marginally and non-marginally bound cases are considered for the
collapsing model. Finally dark energy in the form of anisotropic
fluid is investigated and it is found to be similar to ref. [12].
\end{abstract}


\maketitle

\section{\normalsize\bf{Introduction}}

In recent past, there are remarkable observational evidences
which contradict the present day prediction of Standard
Cosmology. The observed data from the high red shift of type I-a
supernova [1] suggests that the universe at present is
accelerating instead of deceleration (a prediction of Standard
Cosmology). This was confirmed by the observed data from
measurements of the fluctuations in the power spectrum of the
cosmic microwave background radiation [2] and large scale
structure [3].\\

To incorporate such accelerating phase within the frame-work of
Einstein's general relativity one requires source of repulsive
gravity, termed as dark energy. This component of the matter
distribution of the universe should have a large negative
pressure and hence violates the strong energy condition and
dominates over matter at present. Also astronomical observations
predict that at present the universe contains approximately
$\frac{2}{3}$ dark energy and $\frac{1}{3}$ dark matter.
Although, the nature of both dark matter and dark energy are
still unknown, yet based on their characteristics, different
models have been suggested namely tiny positive cosmological
constant, quintessence, phantoms, chaplygin gas, dark energy in
brane-worlds and many others (see the references [4]).\\

The dark energy, by virtue of its repulsive gravitational nature,
is interesting to study gravitational collapse and formation of
black hole. As all massive stars do not form black holes (may be
neutron stars or white dwarfs), so it is generally speculated [5]
that the dark energy may play an important role in the collapsing
stars.\\

In the present study, a collapsing spherical star is considered
having finite thickness. The star is made of dust cloud in the
background of dark energy. Let $\Sigma$ be the boundary of the
star and $V^{+}$ and $V^{-}$ indicate the exterior and interior
region of the star. The spherically symmetric inhomogeneous
space-time (in $V^{-}$) can be described by the
Lemaitre-Tolman-Bondi (LTB) metric element as

\begin{equation}
V^{-}:~~ds_{-}^{2}=-dt^{2}+\frac{R'^{2}}{1-f(r)}dr^{2}+R^{2}d\Omega^{2}
\end{equation}

with $R=R(r,t)$.\\

Due to the choice of the co-moving coordinates, the surface
$\Sigma$ can be identified as

\begin{equation}
\sum:~~r=r_{_{\Sigma}},~~~\text{a constant}.
\end{equation}

and the metric on it is in the form

\begin{equation}
ds_{\Sigma}^{2}=-d\tau^{2}+R_{\Sigma}^{2}(\tau)~d\Omega^{2}
\end{equation}

Thus we have

\begin{equation}
t=\tau ~~\text{and}~~ R_{\Sigma}(\tau)=R(r_{_{\Sigma}},\tau)
\end{equation}

In $V^{+}$, the metric of the space-time in general can be taken
as

\begin{equation}
V^{+}:~~ds_{+}^{2}=-A^{2}(T,z)dT^{2}+B^{2}(T,z)(dz^{2}+z^{2}d\Omega^{2})
\end{equation}

So the bounding surface $\Sigma$ for the coordinates of the
exterior space can be expressed as

\begin{equation}
z=z_{0}(T)
\end{equation}

Then the junction conditions namely
$ds_{-}^{2}|_{\Sigma}=ds_{+}^{2}|_{\Sigma}$ give

\begin{equation}
i)~~\frac{dT}{dt}=\frac{1}{\sqrt{A^{2}-\left(\frac{dz_{0}}{dT}\right)^{2}B^{2}}}
\end{equation}

\begin{equation}
ii)~~R(r_{_{\Sigma}},t)=Z_{0}(T)~B(T,z_{0}(T))
\end{equation}

The other junction conditions (relating to extrinsic curvature)
due to Israel [6] depend on the choice of the space-time outside
the star. If the bounding surface $\Sigma$ is an energy layer i.e,
an infinitely thin matter shell appears on $\Sigma$, then the
extrinsic curvature is not continuous across $\Sigma$ but the jump
discontinuity depends on the matter of the thin shell over
$\Sigma$. On the other hand, the continuity of the extrinsic
curvature components demand that no such shell exists on $\Sigma$.
In fact, whenever the space-time $V_{-}$ is fixed, then the
space-time $V_{+}$ will determine whether this shell appears or
not. In the present paper, it is assumed that no such thin shell
exists on $\Sigma$ and extrinsic curvature is continuous across
$\Sigma$. The paper is organized as follows: In Section II, basic
equations for collapsing inhomogeneous spherical star model are
presented. Sections III and IV deal with marginally bound and
non-marginally bound cases with dark energy in the form of
perfect fluid while anisotropic fluid form is used for dark
energy in Section V. Finally, the paper ends with conclusion in
Section VI.\\

\section{\normalsize\bf{Inhomogeneous spherically symmetric star model
with dark energy in the form of perfect fluid}}

The energy-momentum tensor for the matter field consisting of an
inhomogeneous dust (dark matter) $\rho_{_{DM}}(r,t)$ and
homogeneous dark energy in the form of perfect fluid is given by

\begin{equation}
T_{\mu\nu}=(\rho_{_{DM}}+\rho+p)u_{\mu}u_{\nu}+pg_{\mu\nu}
\end{equation}

where in the comoving representation, four-velocity
$u_{\mu}=\delta_{\mu}^{t}$.\\

The explicit form of the energy conservation equations
$T_{\mu;\nu}^{\nu}=0$ for the metric (1) are

\begin{equation}
\dot{\rho_{_{DM}}}+\left(2\frac{\dot{R}}{R}+\frac{\dot{R}'}{R'}\right)\rho_{_{DM}}=Q(r,t)
\end{equation}

and

\begin{equation}
\dot{\rho}+\left(2\frac{\dot{R}}{R}+\frac{\dot{R}'}{R'}\right)(\rho+p)=-Q(r,t)
\end{equation}

where (~$\dot{}$~) represents differentiation with respect to time
while (~$'$~) represents the same with respect to $r$ and $Q(r,t)$
stands for the interaction between dark matter and dark energy.\\

The non-vanishing components of the Einstein field equations foe
the metric (1) having matter field in the form (9) are given by
(with $\kappa=\frac{8\Pi G}{c^{4}}=1$)

\begin{equation}
\frac{f}{R^{2}}+\frac{f'}{RR'}+\frac{\dot{R}^{2}}{R^{2}}+2\frac{\dot{R}}{R}\frac{\dot{R}'}{R'}=\rho_{_{DM}}+\rho
\end{equation}

\begin{equation}
\frac{f}{R^{2}}+\frac{\dot{R}^{2}}{R^{2}}+2\frac{\ddot{R}}{R}=-p
\end{equation}

\begin{equation}
\frac{f'}{2RR'}+\frac{\dot{R}}{R}\frac{\dot{R}'}{R'}+\frac{\ddot{R}}{R}+\frac{\ddot{R}'}{R'}=-p
\end{equation}

In general the interaction term $Q(r,t)$ is non-zero. But if
$Q(r,t)$ is zero then from the conservation equation (11), $R$
should be in the product form.\\

Since in the present work, one mainly considers gravitational
collapse, so one assumes $\dot{R}<0$.\\
The apparent horizon is characterized by

\begin{equation}
R_{,\alpha}R^{,\alpha}=0
\end{equation}

If $t=t_{ah}(r)$ be the time of formation of apparent horizon,
then from the above equation (15)

\begin{equation}
\dot{R}^{2}\left(r,t_{ah}(r)\right)=1-f(r)
\end{equation}

The mass function at comoving coordinate $'r'$ is given by [7]

\begin{equation}
m(r,t)=\frac{1}{2}R~(1-R_{,\alpha}R^{,\alpha})=\frac{1}{2}R\left(\dot{R}^{2}+f(r)\right)
\end{equation}

So the total mass of the collapsing star at any time $\tau$ is
[8],

\begin{equation}
M(\tau)=\frac{1}{2}R_{\Sigma}(\tau)\left(\dot{R_{\Sigma}}^{2}+f(r)\right)
\end{equation}

Since the collapsing star is assumed to be not trapped initially
(at $\tau=\tau_{i}$), so on the initial hypersurface
$\tau=\tau_{i}$ one should have

\begin{equation}
\dot{R}^{2}(r,\tau_{i})+f(r)-1<0~~.
\end{equation}

\section{\normalsize\bf{Marginally bound case}: $f=0$}

In this case the hypersurfaces $t=$constant have zero curvature.
Now equating the field equations (13) and (14), we have a
differential equation in $R$ which has a first integral

\begin{equation}
2\frac{\ddot{R}}{R}+\frac{\dot{R}^{2}}{R^{2}}=\mu(t)
\end{equation}

Integrating once more, the evolution equation for $R$ is

\begin{equation}
\dot{R}^{2}=\frac{\mu(t)}{3}R^{2}+\frac{g(r)}{R}
\end{equation}

where $\mu(t)$ and $g(r)$ are arbitrary functions of $t$ and $r$
respectively.\\
Now from the field equation (13) using (20), one finds

\begin{equation}
p(t)=-\mu(t)
\end{equation}

As for dark energy $p(t)$ is negative, so the arbitrary function
$\mu(t)$ is positive. Then from the field equation (12) one gets

\begin{equation}
\rho_{_{DM}}+\rho=\mu(t)+\frac{g'(r)}{R^{2}R'}
\end{equation}

It is to be noted that if a barotropic equation of state of the
form $p=\epsilon\rho$ is taken for dark energy, then one arrives
at a contradiction to satisfy the conservation equations (10) and
(11).\\

In particular, if $\mu$ is taken to be constant and the dark
energy is in the form of cosmological constant i.e,
$\rho=-p=\mu$, then from (23)\\

~~~~~~~~~~~~~~~~~~~~~~~~~~~~~~~~~~~~~~~~~~~~~~~~~~~~~~~~~~~~
$~\rho_{_{DM}}=g'(r)/R^{2}R'$\\

and from conservation equations $Q=0$.\\

Thus when the dark energy is in the form of cosmological constant
then the model corresponds to inhomogeneous dust collapse with a
cosmological constant.\\

Now, apparent horizon will form if the cubic equation\\

~~~~~~~~~~~~~~~~~~~~~~~~~~~~~~~~~~~~~~~~~~$\mu R^{3}-3R+3g=0$\\

has at least one positive root. Various cases are as follows:\\

~~~i) For $\mu>0$ and $3g=\frac{2}{\sqrt{\mu}}$, a unique
apparent horizon will form, given by\\

~~~~~~~~~~~~~~~~~~~~~~~~~~~~~~~~~~~~~~~~~~~$R_{ah}(r)=\frac{1}{\sqrt{\mu}}$.\\

~~ii) If $\mu>0$ and $3g<\frac{2}{\sqrt{\mu}}$, then the cubic
equation has two positive real roots, which correspond to
cosmological and black hole horizons given by\\

~~~~~~~~~~~~~~~~~~~~~~~~$R_{c}(r)=\frac{2}{\sqrt{\mu}}Cos\left[\frac{1}{3}Cos^{-1}\left(-\frac{3}{2}g\sqrt{\mu}\right)\right]$\\
and\\

~~~~~~~~~~~~~~~~~~~~~~~~$R_{b}(r)=\frac{2}{\sqrt{\mu}}Cos\left[\frac{4\Pi}{3}+\frac{1}{3}Cos^{-1}\left(-\frac{3}{2}g\sqrt{\mu}\right)\right]$.\\

~iii) For $\mu>0$ and $3g>\frac{2}{\sqrt{\mu}}$ there are no
positive roots and hence there are no apparent horizons.\\

The details of this collapsing process has been studied in
Ref. [13].\\

The effect on the collapsing star will be studied separately for
a dust cloud and dark energy to examine the different roles that
may play during the evolution. Also their joint effects (with
or without interaction) will be investigated subsequently.\\

{\bf Case I}:~~~~~~~~$(\rho_{_{DM}}\neq0,~\rho=0=p)$\\

This case corresponds to inhomogeneous dust collapse in LTB
model. This has been studied exhaustively by Joshi and
collaborators [9] (for higher dimensional study, see Ref. [10]).\\

{\bf Case II}:~~~~~~~$(\rho_{_{DM}}=0,~p=\epsilon\rho\neq 0)$\\

As $Q(r,t)=0$ [from equation (10)], so area radius $R$ must be in
the separable form

\begin{equation}
R(r,t)=\sqrt[3]{g(r)}~R_{t}(t)
\end{equation}

for consistency of the conservation equation (11) and field
equations (12)-(14).\\

Then from the conservation equation
$\rho(t)=\rho_{0}/R_{t}^{3(1+\epsilon)}$\\

and

\begin{equation}
\mu(t)=-\frac{\epsilon\rho_{0}}{R_{t}^{3(1+\epsilon)}}
\end{equation}

Thus the differential equation in $R_{t}$ is

\begin{equation}
\dot{R_{t}}^{2}=-\frac{\epsilon\rho_{0}}{3}R_{t}^{-(3\epsilon+1)}+\frac{1}{R_{t}}
\end{equation}

In the following, this differential equation in $R_{t}$ is solved
for $\epsilon=-1,0,1$ and $-\frac{1}{3}$.\\

{\bf a) Dark energy in the form of cosmological
constant}:    $(\epsilon=-1)$\\

Equation (26) can be solved easily as

\begin{equation}
R_{t}^{\frac{3}{2}}=\sqrt{\frac{3}{\rho_{0}}}Sinh\left[\frac{\sqrt{3\rho_{0}}}{2}(t-t_{0})\right]~~
\text{or}~~\sqrt{\frac{3}{\rho_{0}}}Sinh\left[\frac{\sqrt{3\rho_{0}}}{2}(t_{0}-t)\right]
\end{equation}

according as $\dot{R}>0$ or $<0$.\\

The mass function has the expression
~~~$m(r,t)=\frac{1}{2}g(r)\left(1+\frac{\rho_{0}R_{t}^{3}}{3}\right)$.\\

Hence it represents either an ever expanding model of the
universe starting from the big-bang at $t=t_{0}$ or a collapsing
model of the universe having infinite volume at $t=-\infty$ and
collapse to big crunch at $t=t_{0}$ with a finite mass.\\

{\bf b)} $\epsilon=0$:\\

One can see easily that it corresponds to a pure homogeneous dust
(not dark energy) with $R_{t}\sim t^{\frac{2}{3}}$ as expected in
standard FRW model.\\

{\bf c) Dark energy in the form of perfect fluid:
$(\epsilon=1)$}\\

Here the scale factor grows with time as

\begin{equation}
R_{t}^{3}=\frac{\rho_{0}}{3}+\frac{9}{4}(t-t_{0})^{2}
\end{equation}

with $\rho(t)=\rho_{0}/R_{t}^{6}$.\\

{\bf d)} $\epsilon=-\frac{1}{3}$:\\

In this case we have

\begin{equation}
\sqrt{R_{t}^{2}\rho_{0}^{2}+R_{t}\rho_{0}}-Sinh^{-1}\left(\sqrt{R_{t}\rho_{0}}\right)
=\frac{\rho_{0}^{\frac{3}{2}}}{3}(t_{0}-t)
\end{equation}

The mass function has the expression
~~~$m(r,t)=\frac{1}{2}g(r)\left(1+\frac{\rho_{0}R_{t}}{9}\right)$.\\

{\bf Case III}:~~~~~~~$(\rho\neq0,~\rho_{_{DM}}\neq0,~Q=0)$\\

As $Q=0$, so similar to Case II, $R$ must be in the product
separable form (i.e, eq. (24)) for consistency of the
conservation equation (10). The expressions for the matter
densities become

\begin{equation}
\rho_{_{DM}}=h(r)R_{t}^{-3}~~~\text{and}~~~\rho=\rho_{0}R_{t}^{-3(1+\epsilon)}
\end{equation}

with $\rho_{0}$, an arbitrary constant and $h(r)$, an arbitrary
function of $r$ alone. Also the evolution equation for $R_{t}$
takes the form of equation (26).\\

The consistency of the field equation (12) demands $\rho_{_{DM}}$
to be also homogeneous with $h(r)=3$. Thus it is very similar to
case II and physical consequences are identical. Note that here
the role of dark matter is insignificant.\\

{\bf Case IV}:~~~~~~$(\rho\neq0,~\rho_{_{DM}}\neq0,~Q\neq 0)$\\

For the interacting dark matter and dark energy, the evolution
equation for $R$, the expressions for $p(t)$ and ($\rho(t)+
\rho_{_{DM}}$) are given by equations (21)-(23) respectively.\\

Now for explicit solution of $R$ one assumes (see Ref. [11])

\begin{equation}
\mu(t)=\mu_{0}t^{-s},~\text{($\mu_{0}$ and s are positive
constants)}
\end{equation}

and the general solution for $R$ has the form [12]

\begin{equation}
R^{\frac{3}{2}}=\left\{
\begin{array}{ll}
\sqrt{t}\left\{C_{1}J_{\xi}\left[\frac{2\sqrt{\lambda}}{|s-2|}t^{-\frac{s-2}{2}}\right]
+C_{2}Y_{\xi}\left[\frac{2\sqrt{\lambda}}{|s-2|}t^{-\frac{s-2}{2}}\right]\right\}\\\\
\sqrt{t}\left\{C_{1}J_{\xi}\left[\frac{2\sqrt{\lambda}}{|s-2|}t^{-\frac{s-2}{2}}\right]
+C_{2}J_{-\xi}\left[\frac{2\sqrt{\lambda}}{|s-2|}t^{-\frac{s-2}{2}}\right]\right\}\\\\
C_{1}t^{q_{1}}+C_{2}t^{1-q_{1}}
\end{array}
\right.
\end{equation}

according as $\xi$ is an integer, non-integer and $s=2$. Hence
$C_{1}$ and $C_{2}$ are arbitrary functions of $r$ and\\

$\xi=\frac{1}{s-2}$, $\lambda=\frac{3\mu_{0}}{4}$,
$q_{1}=\frac{1}{2}\left(1+\sqrt{1-3\mu_{0}}\right)$.\\\\

It can be shown (for details see Ref. [12]) that the model
approaches isotropy along the fluid world line as
$t\rightarrow\infty$.\\

\section{\normalsize\bf{Non-marginally bound case}: $f\neq0$}

In this case, the first integral of equation (13) gives the
dynamical equation for the area radius as

\begin{equation}
\dot{R}^{2}=-\frac{p(t)R^{2}}{3}+\frac{g(r)}{R}-f(r)
\end{equation}

with $g(r)$ as arbitrary function of $r$. The explicit form of
total energy density is (obtained from equation (12))

\begin{equation}
\rho+\rho_{_{DM}}=\frac{g'(r)}{R^{2}R'}-p(t)
\end{equation}

In particular, if one chooses $\rho(t)=-p(t)$, then

\begin{equation}
\rho_{_{DM}}=g'(r)/R^{2}R'
\end{equation}

and consequently from the conservation equations (10) and (11)
one obtains respectively\\

~~~~~~~~~~~~~~~~~~~~~~~~~~~~~~~~~~~~~~~~~~~~~~~$Q=0$~~~~ and
~~~~$\rho=$constant $\lambda$ (say).\\

Hence as in marginally bound case, this can also corresponds to a
collapsing dust (inhomogeneous) with a cosmological constant.
Note that, due to the difference in the evolution equations [see
eqns. (33) and (21)], the collapsing processes will not be
identical.\\

One may note that if the separable (product) form of $R$ is
assumed, i.e, $R=\sqrt{f(r)}~\xi(t)$, then $\xi(t)$ satisfies

\begin{equation}
\dot{\xi}^{2}+1=-\frac{p(t)\xi^{2}}{3}+\frac{\eta(t)}{\xi}
\end{equation}

where $\eta(t)$ is an arbitrary function of $t$ alone. The form
of energy density becomes

\begin{equation}
\rho+\rho_{_{DM}}=\frac{3\eta(t)}{\xi^{3}}-p(t)
\end{equation}

In particular if $\rho(t)=-p(t)=\lambda(t)$, then

\begin{equation}
\rho_{_{DM}}=\frac{3\eta(t)}{\xi^{3}}
\end{equation}

and the arbitrary function $\eta(t)$ is related to $\lambda(t)$
by the relation

\begin{equation}
\dot{\eta(t)}=-\frac{\dot{\lambda(t)}\xi^{3}}{3}
\end{equation}

Thus dark energy behaves as time dependent cosmological constant
and the model describes homogeneous dust collapse with time
dependent cosmological term.\\

\section{\normalsize\bf{Dark energy in the form of anisotropic
fluid}}

Here if the dark energy is in the form of anisotropic fluid with
energy momentum tensor

\begin{equation}
T_{(DE)\mu}^{\nu}=diag(-\rho,p_{r},p_{_{T}},p_{_{T}})
\end{equation}

then the total energy momentum tensor is

\begin{equation}
T_{(T)\mu}^{\nu}=\rho_{_{DM}}u_{\mu}u^{\nu}+T_{(DE)\mu}^{\nu}
\end{equation}

So the conservation equation gives

\begin{equation}
\left.
\begin{array}{cc}
\dot{\rho_{_{DM}}}+\left(2\frac{\dot{R}}{R}+\frac{\dot{R}'}{R'}\right)
\rho_{_{DM}}=Q(r,t)\\\\
\dot{\rho}+\left(2\frac{\dot{R}}{R}+\frac{\dot{R}'}{R'}\right)\rho+
\left(\frac{\dot{R}'}{R'}p_{r}+\frac{2\dot{R}}{R}p_{_{T}}\right)=-Q(r,t)
\end{array}
\right\}
\end{equation}

and

\begin{equation}
p_{r}'+2(p_{r}-p_{_{T}})\frac{R'}{R}=0
\end{equation}

One may note that the dark energy in the form of anisotropic
fluid must be inhomogeneous, otherwise from the conservation
equation (43), $p_{r}'=0$ implies $p_{r}=p_{_{T}}$ i.e, fluid is
isotropic (discussed in the previous sections). Conversely, if
the dark energy fluid is inhomogeneous (i.e, $p_{r}'\neq0$), then
again from equation (43), $p_{r}\neq p_{_{T}}$, i.e, the
fluid must be anisotropic.\\

Now from the field equation (13) (replacing $p$ by $p_{r}$), the
evolution equation for $R$ becomes (integrating once),

\begin{equation}
\dot{R}^{2}=\frac{g(r)}{R}-\frac{1}{R}\int p_{r}(r,t)R^{2}dR-f(r)
\end{equation}

Now assuming the regularity of the initial radial pressure at the
centre and blowing up at the singularity, the form of $p_{r}$ can
be taken as (for details see Ref. [12])

\begin{equation}
p_{r}=\frac{\alpha(r)}{R^{n}}
\end{equation}

where $\alpha(r)$ is an arbitrary function of $r$ such that
$\alpha(r)\sim r^{n}$ near $r=0$. and $n(>0)$ is any constant.\\

Thus the radial velocity of the collapsing shells at a distance
$r$ from the centre is given by

\begin{equation}
\dot{R}^{2}=\frac{g(r)}{R}-\frac{\alpha(r)}{3-n}R^{2-n}-f(r)
\end{equation}

Also from the conservation equations, the expressions for the
energy densities become

\begin{equation}
\rho_{_{DM}}=\frac{\rho_{0}(r)}{R^{2}R'}+\frac{1}{R^{2}R'}\int
QR^{2}R'dt
\end{equation}
and
\begin{equation}
\rho=\frac{H'(R,t)}{R^{2}R'}-\frac{1}{R^{2}R'}\int QR^{2}R'dt
\end{equation}

where $H(R,t)=\rho_{1}(r)-\frac{\alpha(r)}{3-n}R^{3-n}$ and the
arbitrary functions $\rho_{0}$ and $\rho_{1}$ are restricted by
the relation [consistency of equation (12)]

\begin{equation}
\rho_{1}'+\rho_{0}=g(r)
\end{equation}

The expression for tangential stress becomes

\begin{equation}
p_{_{T}}=\frac{\alpha(r)}{R^{n}}\left[1-\frac{nR'}{2(R'+R\nu')}\right]
+\frac{\alpha'(r)}{2R^{n-1}(R'+R\nu')}~,~~~(n\neq3)
\end{equation}

A similar collapsing process has been studied extensively in Ref.
[12] for marginally bound case only ($f=0$) and it is calculated
that in general, pressure tries to resist the formation of naked
singularity. Thus formation of a black hole from the collapsing
star due to the presence of dark energy in the form of
anisotropic fluid is more favorable than formation of a naked
singularity.\\

\section{\normalsize\bf{Conclusion}}

The paper deals with a detail study of an inhomogeneous
spherically symmetric star model having dark matter in the
background of dark energy as the matter content. The dark matter
is taken in the form of inhomogeneous dust while for dark energy,
both homogeneous perfect fluid and anisotropic fluid models are
considered separately. One may note that the present study can
easily be extended to quasi-spherical collapsing star models and
the results will be identical.\\

For dark energy in the form of perfect fluid, both marginally
bound ($f=0$) and non-marginally bound ($f\neq0$) cases are
considered in two different sections. When $f=0$, dark matter and
dark energy are considered both separately and in combined form.
It is observed that due to the presence of dark energy, trapped
surfaces do not form at all and collapsing models lead to big
crunch singularity. In a particular situation (when
$\epsilon=-1$), the non-marginally bound case represents
inhomogeneous dust collapse with a cosmological constant while
assuming separable form of area radius, the model corresponds to
homogeneous dust collapse with time dependent cosmological term.
As for inhomogeneous dust with cosmological constant, trapped
surfaces may be possible while for homogeneous dust with time
dependent cosmological term, formation of trapped surfaces is not
possible. Hence naked singularity is more favorable in second
case than in first case.\\

Lastly, when anisotropic fluid represents dark energy, it is
found that black hole formation is more favorable than naked
singularity. Therefore, one may conclude that inhomogeneity, both
in dark matter and dark energy favors formation of trapped
surfaces, while homogeneous cases support the singularity to be
naked.\\

{\bf References}:\\
\\
$[1]$ A.G.Riess et al, Supernova Search Team Collaboration, {\it
Astrophys.J.} {\bf 607}, 665 (2004); {\it Astron.J.} {\bf 116},
1009 (1998); S.Perlmutter et al, Supernova Cosmology Project
Collaboration, {\it Astrophys.J.} {\bf 517}, 565 (1999).\\
$[2]$ D.N. Spergel et al, {\it Astrophys.J.Suppl.} {\bf 148}, 175
(2003); J.L. Sievers et al, {\it Astrophys.J.} {\bf 591} 599
(2003); R.Stompor et al, {\it Astrophys.J.} {\bf 561}, L7
(2001).\\
$[3]$ M.Tegmark et al, {\it Phys.Rev.D} {\bf 69}, 103501 (2004);
E.Hawkings et al, {\it Mon.Not.Roy.Astron.Soc} {\bf 346}, 78
(2003); K.Abazajian et al [SDSS Collaboration], {\it Astron.J.}
{\bf 128}, 502 (2004); {\it Astron.J.} {\bf 126},
2081 (2003).\\
$[4]$ T.Padmanabhan, {\it Phys.Rep.} {\bf 380} 235 (2003);
P.J.E.Peebles and B.Ratra, {\it Rev.Mod.Phys.} {\bf 75}, 559
(2002); S.M Carroll, {\it Living Rev.Rel.} {\bf 4}, 1 (2001);
V.Sahni and A.A.Starobinsky, {\it IJMPD} {\bf 9}, 373 (2000);
V.Sahni, "Cosmological Surprises from Braneworld Models of Dark
Energy", {\it arxiv.astro-ph/}0502032 (2005).\\
$[5]$ http://www.nature.com/news/2005/050328/full/050328-8 html;
http://www.bioon.com/TILS/news/200504\\
/97002.html; See also, D.F.Mota, C.Van de Bruck, {\it
Astron.Astrophys.} {\bf 421}, 71 (2004); P.G.Ferreira and
M.Joyce, {\it Phys.Rev.Lett.} {\bf 79}, 4740 (1997).\\
$[6]$ W.Israel, {\it Nuovo Cim.} {\bf 44B}, 1 (1966); ibid, {\bf
48B}, 463 (1967).\\
$[7]$ M.E.Cahill and G.C.McVittie, {\it J.Math.Phys.} {\bf 11},
1382 (1970).\\
$[8]$ Rong-Gen Cai and A.Wang, "Black hole formation from
collapsing dust fluid in the background of dark energy", (2006);
{\it astro-ph/}0505136.\\
$[9]$ P.S.Joshi, "Global Aspects in Gravitation and Cosmology",
(1993), Oxford University Press, Oxford; P.S.Joshi and
I.H.Dwivedi, {\it Commun.Math.Phys.} {\bf 166} 117 (1994);
{\it Class.Quant.Grav.} {\bf 16} 41 (1999).\\
$[10]$ A.Banerjee, U.Debnath and S.Chakraborty {\it Int.J.Mod.Phys.D}
{\bf 12} 1255 (2003); U.Debnath and S.Chakraborty, {\it J.of Cosmology
and Astroparticle Physics} {\bf 5} 001 (2004).\\
$[11]$ D.A.Szafron, {\it J.Math.Phys.} {\bf 18}, 1673 (1977);
D.A.Szafron and J.Wainwright, {\it J.Math.Phys.} {\bf 18}, 1668 (1977).\\
$[12]$ Subenoy Chakraborty, S.Chakraborty and U.Debnath, {\it
Int.J.Mod.Phys.D} {\bf 14} 1707 (2005).\\
$[13]$ S.S.Deshingkar, S.Jhingan, A.Chamorro and P.S.Joshi, {\it
Phys.Rev.D} {\bf 63}, 124005 (2001); M.Cissoko, J.C.Fabris,
J.Gaviel, G.L.Denmat and N.O.Santos, {\it gr-qc/}9809057;
D.Markovic and S.L.Shapiro, {\it Phys.Rev.D} {\bf 61}, 084029
(2000); K.Lake, {\it Phys.Rev.D} {\bf 62}, 027301 (2000); U.Debnath,
S.Nath and S.Chakraborty, {\it Mon.Not.R.Astr.Soc.} (2006) ({\it accepted}). \\

\end{document}